\documentclass[12pt]{iopart}
\usepackage{color}
\usepackage{graphicx}
\usepackage{epsfig}
\usepackage[T1]{fontenc}
\include{epsf}

\begin{document}

\title[Entropy production for asymmetric diffusion of particles]{Entropy production for asymmetric diffusion of particles}

\author{M. O. Hase$^1$, T. Tom\'e$^2$ and M. J. de Oliveira$^2$}

\address{$^1$ Escola de Artes, Ci\^encias e Humanidades,
Universidade de S\~ao Paulo, 
Avenida Arlindo B\'ettio, 1000,
03828-000 S\~ao Paulo, S\~ao Paulo, Brazil}

\address{$^2$ Instituto de F\'{\i}sica,
Universidade de S\~ao Paulo, 
Caixa Postal 66318,
05314-970 S\~ao Paulo, S\~ao Paulo, Brazil}

\ead{mhase@usp.br}

\begin{abstract}
We analyse a non-equilibrium exclusion process in which particles are created and annihilated in pairs and hop to the the right or to the left with different transition rates, $p$ and $q$, respectively. We have studied the dynamics of a single particle, and exactly determined the entropy, entropy production rate and entropy flux as functions of time. In the system of many particles, we have characterised the system by its probability distribution, as well as the entropy production rate in close forms, provided that $p+q$ equals the sum of dimers creation and annihilation rates. The general case, where this constraint is absent, was considered at pair approximation level; the time-dependent behaviour of the system was analysed, and the stationary entropy production was determined. In all cases, in the stationary regime, we showed that the entropy production rate is a bilinear form in the current of particles and the force $\ln(p/q)$.
\end{abstract}

\pacs{05.70.Ln, 02.50.Ga, 05.50.+q}

\maketitle

%----------------------------------------------------------
\section{Introduction}

Non-equilibrium statistical physics presents fundamental questions that may require the use of approaches and quantities which has no analogue in equilibrium context. One of these quantities is the production of entropy, which vanishes in equilibrium and is positive in non-equilibrium and may thus characterize a system out of thermodynamic equilibrium. The time derivative of entropy $S$ of a system can be decomposed into two components \cite{NP77} as
\begin{eqnarray}
\frac{\textnormal{d}S}{\textnormal{d}t} = \Pi - \Phi,
\label{1}
\end{eqnarray}
where one has the entropy production rate, $\Pi$, and entropy flux, $\Phi$. The former is the contribution to the entropy by the system itself and is always non-negative. The latter, the entropy flux, is the contribution due to the environment, and can have either signs: the flux is positive if the entropy leaves the system. These functions distinguish different dynamics regimes, since in the stationary state they are equal, $\Pi=\Phi$ (but not necessarily zero); furthermore, in equilibrium state, they both vanish, $\Pi=\Phi=0$.

Equation (\ref{1}) is useful in the study of models in statistical mechanics as long as we know the statistical definitions of the quantities $S$, $\Pi$ and $\Phi$ for systems out of equilibrium. Here we are interested in models described by a continuous time Markovian process, that is, governed by the master equation \cite{vK81,TO15}
\begin{eqnarray}
\frac{\textnormal{d}}{\textnormal{d}t}P_{i}(t) = \sum_{j} \Big[ W_{ij}P_{j}(t) - W_{ji}P_{i}(t) \Big],
\label{4}
\end{eqnarray}
where $P_{i}(t)$ is the probability of finding the system in state $i$ at time $t$ and $W_{ij}$ is the transition rate from state $j$ to state $i$. The definition of entropy $S$ in thermodynamic equilibrium or out of equilibrium is given by the Boltzmann-Gibbs expression
\begin{eqnarray}
S(t) = - \sum_{i} P_{i}(t)\ln P_{i}(t),
\label{2}
\end{eqnarray}
and the entropy production rate is given by \cite{S76}
\begin{eqnarray}
\Pi(t) = \frac12 \sum_{ij}\Big[W_{ij}P_{j}(t) - W_{ji}P_{i}(t)\Big] \ln\frac{W_{ij}P_{j}(t)}{W_{ji}P_{i}(t)},
\label{5}
\end{eqnarray}
an expression that has been considered by several authors\cite{J84,M86,CT05,Z06,A06,S07,Z07,E09,TO10,O11,E12,HO12,TO12}.

From the time derivative of entropy,
\begin{eqnarray}
\frac{\textnormal{d}S}{\textnormal{d}t} = \frac12 \sum_{ij}\Big[W_{ij}P_{j}(t) - W_{ji}P_{i}(t)\Big] \ln\frac{P_{j}(t)}{P_{i}(t)},
\label{7}
\end{eqnarray}
obtained by the use of the master equation, and the definition of $\Pi$, we find from (\ref{1}) that the entropy flux is given by
\begin{eqnarray}
\Phi(t) = \frac12 \sum_{ij}\Big[W_{ij}P_{j}(t) - W_{ji}P_{i}(t)\Big] \ln\frac{W_{ij}}{W_{ji}}.
\label{6}
\end{eqnarray}
It is seen that the entropy production rate given by expression (\ref{5}) is non-negative. Moreover, it vanishes in thermodynamic equilibrium, in which case detailed balance condition is fulfilled, that is, $W_{ij}P_{j} = W_{ji}P_{i}$. The expression analogous to (\ref{5}) has also been obtained for systems described by a Fokker-Planck equation \cite{T06}.

The calculation of the production of entropy of lattice models has been done in several models; in some cases, by the use of numerical simulations \cite{CT05,O11,HO12,TO12}. In this paper we consider a system of particles moving along a one-dimensional lattice with periodic boundary conditions. They hop to the right or to the left with distinct rates denoted by $p$ and $q$, respectively, so that in the stationary state the system is not in thermodynamic equilibrium and the entropy production rate is non-zero. We analyse two models: the first is a system of non-interacting particles in which case the problem is reduced to solving the master equation for just one-particle that performs an asymmetric random walk along a chain of $L$ sites. The second model is a system of exclusion particles that hop to the right and left with transition rates $p$ and $q$ along a chain of $L$ sites. In addition, they may be created in pairs, with rate $c$, and annihilated in pairs, with rate $a$. Both models are solved exactly, the second with the condition $a+c=p+q$, but we have also analysed the case without this constraint in pair approximation. From the solutions, we have determined the entropy production rate. In the stationary state we show that it can be written in the bilinear form \cite{TO12}
\begin{eqnarray}
\Pi = L J X,
\end{eqnarray}
where $X=\ln(p/q)$ and $J$ is the current of particles which is shown to vanish when $p\to q$ as $J\sim X$. Therefore, near thermodynamic equilibrium, which occurs when $p=q$, the entropy production rate vanishes as $\Pi\sim X^2$.

%----------------------------------------------------------
\section{One-particle system}

\subsection{Discrete formulation}

A one-dimensional diffusive particle will be considered here. The state of this particle can be identified with the (discrete) position $n$ it can occupy over a chain. The distribution probability $P_{n}$ is governed by the master equation,
\begin{eqnarray}
\frac{\textnormal{d}}{\textnormal{d}t}P_{n}(t) = \sum_{m}\Big[ W_{nm}P_{m}(t) - W_{mn}P_{n}(t) \Big],
\label{me}
\end{eqnarray}
where $W_{nm}$ is the transition rate from state $m$ to $n$, given by
\begin{eqnarray}
W_{nm} = \left\{
\begin{array}{lcl}
p, & & n-m = 1, \\
q, & & n-m=-1, \\
0, & & |n-m|>1.
\end{array}
\right.
\label{W}
\end{eqnarray}
In this setup, the particle jumps to the right with rate $p$ and to the left with rate $q$. The parameters $p$ and $q$ will be taken to be both strictly positive. The transition is short-ranged in the sense that the particle moves to one of its neighbour site per movement. The master equation for this model can be restated as
\begin{eqnarray}
\frac{\textnormal{d}}{\textnormal{d}t}P_{n}(t) = pP_{n-1}(t) + qP_{n+1}(t) - \left(p+q\right)P_{n}(t)\,,
\label{me_dif}
\end{eqnarray}
which can be solved exactly. If the initial condition is taken to be $P_{n}(0)=\delta_{n0}$, the probability distribution is
\begin{eqnarray}
P_{n}(t) = e^{-(p+q)t}\left(\frac{p}{q}\right)^{n/2} I_{n}(2t\sqrt{pq}),
\label{Pn}
\end{eqnarray}
where $I_{n}(z)$ stands for modified Bessel function (first kind). 

We start by calculating the entropy flux. Applying formula (\ref{6}) to the present case, one has
\begin{eqnarray}
\Phi = \frac{1}{2}\sum_{n}\Big[ pP_{n}(t) - qP_{n+1}(t) \Big]\ln\frac{p}{q} +\frac{1}{2}\sum_{n}\Big[ q P_{n}(t) - pP_{n-1}(t)\Big]\ln\frac{q}{p},
\label{Phi}
\end{eqnarray}
and it is straightforward to see that 
\begin{eqnarray}
\Phi = (p-q)\ln\frac{p}{q}.
\label{Phi_P}
\end{eqnarray}
Note that this value for the entropy flux does not depend on time and is a non-negative quantity.

Instead of using formula (\ref{5}) to calculate the entropy production rate, we determine the entropy derivative $\textnormal{d}S/\textnormal{d}t$ and sum to $\Phi$ to get $\Pi=\Phi+\textnormal{d}S/\textnormal{d}t$. Now from formula (\ref{7}) we have
\begin{eqnarray}
\frac{\textnormal{d}S}{\textnormal{d}t} = \sum_{n}\Big[pP_{n}(t)-qP_{n+1}(t)\Big] \ln\frac{P_{n}(t)}{P_{n+1}(t)}.
\label{dsdt}
\end{eqnarray}
Replacing (\ref{Pn}) into (\ref{dsdt}) we may calculate $\textnormal{d}S/\textnormal{d}t$. The graphs of $\textnormal{d}S/\textnormal{d}t$ calculated numerically from (\ref{dsdt}) together with $\Phi$ given by (\ref{Phi_P}) and $\Pi=\Phi+\textnormal{d}S/\textnormal{d}t$ are shown in figure 1, where one can see that the entropy production has the tendency to match the entropy flux with time, which leads the time derivative of entropy, $\textnormal{d}S/\textnormal{d}t$, to zero, as expected. The system converges to a stationary state, and the way it reaches is discussed in the next section, where a continuous version of the model is analysed by means of a Fokker-Planck equation.

%-------------
\begin{center}
\begin{figure}
\begin{center}
\includegraphics[width=300pt, height=200pt]{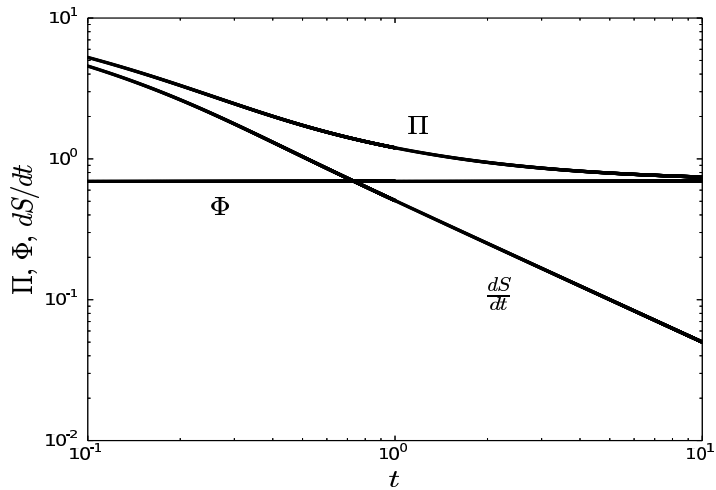}
\end{center}
\caption{Log-log plot of the entropy production rate $\Pi$, entropy flux $\Phi$ and time variation of entropy $\textnormal{d}S/\textnormal{d}t$ versus time. In this example, $p=2$ and $q=1$.}
\end{figure}
\end{center}
%-----------

In the steady state $\Pi=\Phi$ so that from (\ref{Phi_P}) we find the result for the entropy production rate
\begin{eqnarray}
\Pi = (p-q)\ln\frac{p}{q}.
\end{eqnarray}
Now the current of particles is $J=(p-q)/L$ and we may write
\begin{eqnarray}
\Pi = LJX,
\end{eqnarray}
where $X=\ln({p}/{q})$.

\subsection{Continuous formulation}

In the continuous case, which means a (biased) random walk on the real line, one can pass from the master equation (\ref{me_dif}) to a Fokker-Planck equation. The continuous limit of the master equation is obtained as follows. We assume that the possible positions of the particle are $x=bn$ where $b$ is the spacing between consecutive positions. Writing $P_{n}(t) = b P(x,t)$ and expanding $P_{n+1}(t)=bP(x+b,t)$ and $P_{n-1}(t)=bP(x-b,t)$ up to quadratic terms in $b$, we get
\begin{eqnarray}
\frac{\partial}{\partial t}P(x, t) = -\gamma\frac{\partial}{\partial x}P(x, t) + \Gamma\frac{\partial^{2}}{\partial x^{2}}P(x, t),
\label{fp}
\end{eqnarray}
where $\gamma=(p-q)b$ and $\Gamma=(p+q)b^2/2$. The parameter $\Gamma$ is the diffusion constant and $\gamma$ controls the drift of the particle to the right if $\gamma>0$ or to the left if $\gamma<0$. Therefore, $\gamma=0$ corresponds to $p=q$ in the previous discrete model and $p>q$ to positive values of $\gamma$. 

Defining the function
\begin{eqnarray}
J(x, t) = \gamma P(x, t) - \Gamma\frac{\partial}{\partial x}P(x, t),
\label{J}
\end{eqnarray}
the Fokker-Planck equation can be casted as
\begin{eqnarray}
\frac{\partial}{\partial t}P(x, t) = -\frac{\partial}{\partial x}J(x, t),
\label{continuity}
\end{eqnarray}
which is a continuity equation and $J$ can be interpreted as a probability current. In this formulation, the entropy production is given by \cite{T06}
\begin{eqnarray}
\Pi = \frac{1}{\Gamma}\int\textnormal{d}x\,\frac{\left[J(x, t)\right]^{2}}{P(x, t)},
\label{Pi_fp}
\end{eqnarray}
while the flux is
\begin{eqnarray}
\Phi = \frac{\gamma}{\Gamma}\int\textnormal{d}x\,J(x, t).
\label{Phi_fp}
\end{eqnarray}
As usual, the entropy is given by
\begin{eqnarray}
S = -\int\textnormal{d}x\, P(x,t)\ln P(x,t).
\end{eqnarray}

The Fokker-Planck equation (\ref{fp}) can be solved by standard methods, and the probability distribution for the initial condition $P(x,0)=\delta(x)$ is
\begin{eqnarray}
P(x, t) = \frac{1}{\sqrt{4\pi\Gamma t}}\exp\left[-\frac{\left(x-\gamma t\right)^{2}}{4\Gamma t}\right].
\label{P_fp}
\end{eqnarray}
The evaluation of entropy from the probability distribution, given by (\ref{P_fp}), yields
\begin{eqnarray}
S = \frac{1}{2}\ln\left(\Gamma t\right) + \frac{1}{2}\ln\left(4\pi e\right).
\label{S_fp}
\end{eqnarray}
Note that the time derivative of the entropy is
\begin{eqnarray}
\frac{\textnormal{d}S}{\textnormal{d}t} = \frac{1}{2t},
\label{dSdt_fp}
\end{eqnarray}
which decreases algebraically with time. 

Inserting the probability distribution (\ref{P_fp}) into (\ref{Pi_fp}) and (\ref{Phi_fp}) gives, respectively, the entropy production rate,
\begin{eqnarray}
\Pi = \frac{\gamma^{2}}{\Gamma} + \frac{1}{2t},
\end{eqnarray}
the entropy flux,
\begin{eqnarray}
\Phi = \frac{\gamma^{2}}{\Gamma},
\end{eqnarray}
and shows that the entropy flux is independent of time as in the discrete case, and vanishes only if $\gamma=0$. Notice that $dS/dt=\Pi-\Phi$ as it should. In the stationary state, $\Pi=\gamma^2/\Gamma$ or yet $\Pi=2(p-q)^2/(p+q)$.

%----------------------------------------------------------
\section{Many-particles system}

\subsection{Formulation of the problem}
\label{problem}

The analysis above is extended to a system of many diffusive particles with excluded volume interactions. Consider a chain with $L$ sites, each one being occupied by a particle ($\bullet$) or empty ($\circ$). The possible transitions with the respective rates are as follows:
\begin{equation*}
\begin{array}{cccl}
{\bullet}\circ&\stackrel{p}{\longrightarrow}&{\circ}\bullet&
\qquad{\rm diffusion-to-the-right,} \\
{\circ}\bullet&\stackrel{q}{\longrightarrow}&{\bullet}\circ&
\qquad{\rm diffusion-to-the-left,} \\
{\circ}\circ&\stackrel{c}{\longrightarrow}&{\bullet}\bullet&
\qquad{\rm creation,} \\
{\bullet}\bullet&\stackrel{a}{\longrightarrow}&{\circ}\circ&
\qquad{\rm annihilation.}
\end{array}
\end{equation*}
To each site $n$ we associate a variable $s_{n}$ that takes the value $s_{n}=+1$ if the site is occupied ($\bullet$) and the values $s_{n}=-1$ if the site is empty ($\circ$). In terms of these variables, the possible transitions are
\begin{equation*}
\begin{array}{ccc}
(+1,-1) &\stackrel{p}{\longrightarrow}& (-1,+1)  \\
(-1,+1) &\stackrel{q}{\longrightarrow}& (+1,-1)  \\
(-1,-1) &\stackrel{c}{\longrightarrow}& (+1,+1)  \\
(+1,+1) &\stackrel{a}{\longrightarrow}& (-1,-1) 
\end{array}
\end{equation*}
From top to bottom in the scheme above, the particle jumps to right with rate $p$, and hops to left with rate $q$. Moreover, a pair of particles is created and annihilated with rate $c$ and $a$, respectively. Assuming a single transition per unit time, the transition rate $w_n(s)$ from $(s_n,s_{n+1})$ to $(-s_n,-s_{n+1})$, where $s=(s_1,\ldots,s_L)$, can be expressed as
\begin{eqnarray}
w_n(s) = A+Bs_n+Cs_{n+1}+Ds_ns_{n+1},
\label{w}
\end{eqnarray}
where
\begin{eqnarray}
A=\frac{1}{4}(p+q+c+a),
\end{eqnarray}
\begin{eqnarray}
B=\frac{1}{4}(p-q-c+a),
\end{eqnarray}
\begin{eqnarray}
C=\frac{1}{4}(-p+q-c+a),
\end{eqnarray}
\begin{eqnarray}
D=\frac{1}{4}(-p-q+c+a).
\end{eqnarray}

The time evolution of probability distribution is given by the master equation
\begin{eqnarray}
\frac{\textnormal{d}}{\textnormal{d}t}P(s, t) = \sum_{n=1}^{L}\Big[ w_{n}(s^{n, n+1})P(s^{n, n+1}, t) - w_{n}(s)P(s, t) \Big],
\label{me_ff0}
\end{eqnarray}
where $s^{n,n+1}$ denotes the vector $s$ with the variables $s_{n}$ and $s_{n+1}$ replaced by $-s_{n}$ and $-s_{n+1}$, respectively.

\subsection{Stationary solution}

The model defined by the transition rate (\ref{w}) becomes simpler when the condition $D=0$ is imposed because in this case the transition rate $w_n$ becomes linear is the $s_n$ and $s_{n+1}$, that is,
\begin{eqnarray}
w_n(s) = A+Bs_n+Cs_{n+1}.
\label{ws}
\end{eqnarray}
The condition $D=0$ is equivalent to the relation
\begin{eqnarray}
a+c=p+q,
\label{acpq}
\end{eqnarray}
which means that the sum of the creation and annihilation rates are equal to the sum of the diffusion to the right and to the left.

The stationary probability distribution can be shown to be of the form \cite{MO98, O99}
\begin{eqnarray}
P(s) = \frac1Z e^{h(s_1+s_2+\ldots+s_L)},
\label{17}
\end{eqnarray}
where $Z=(2\cosh h)^L$ and $h$ is a parameter do be determined. To this end, we replace (\ref{17}) into the stationary master equation
\begin{eqnarray}
\sum_{n=1}^{L}\Big[ w_{n}(s^{n, n+1})P(s^{n,n+1}) - w_{n}(s)P(s) \Big]=0,
\end{eqnarray}
where $w_n(s)$ is given by (\ref{w}). This equation is to be solved with periodic boundary condition. After substitution we get
\begin{eqnarray}
\sum_{n=1}^{L}\Big[(A-Bs_n-Cs_{n+1})e^{-2h(s_n+s_{n+1})}-(A+Bs_n+Cs_{n+1})\Big]=0.
\end{eqnarray}
Defining $u=\cosh 2h$ and $v=\sinh 2h$, and using periodic boundary condition we may write
\begin{eqnarray}
\sum_{n=1}^L (Av+Bu+Cu)\left[ v(1+s_ns_{n+1}) -2u s_n \right] = 0.
\end{eqnarray}
Therefore, the distribution (\ref{17}) is the solution of the stationary master equation if $Av+Bu+Cu=0$, that is, if
\begin{eqnarray}
\frac{v}{u} = \frac{-B-C}{A} = \frac{c-a}{c+a},
\end{eqnarray}
or $\tanh 2h=(c-a)/(c+a)$, which determines the parameter $h$ in terms of the rates $c$ and $a$. 

From the stationary distribution we may find the magnetisation $m=\langle s_n\rangle=\tanh h$, which is
\begin{eqnarray}
m = \frac{\sqrt{c}-\sqrt{a}}{\sqrt{c}+\sqrt{a}}.
\label{mag}
\end{eqnarray}
We may also determine the pair correlation function $r=\langle s_ns_{n+1}\rangle=\langle s_n\rangle\langle s_{n+1}\rangle=m^2$. Notice that the density of particles is $\rho=P(+)$, or
\begin{eqnarray}
\rho = \frac{\sqrt{c}}{\sqrt{c}+\sqrt{a}}.
\label{dens}
\end{eqnarray}

\subsection{Flux and production of entropy}

The entropy flux, defined by (\ref{Phi}), of a system that has a dynamics governed by the rule (\ref{w}) is given by
\begin{eqnarray}
\Phi = \sum_s\sum_{n}w_{n}(s)P(s)\ln\frac{w_{n}(s)}{w_{n}(s^{n, n+1})},
\label{Phi_w}
\end{eqnarray}
which can be written as the average
\begin{eqnarray}
\Phi = \sum_{n}\left\langle w_{n}(s)\ln\frac{w_{n}(s)}{w_{n}(s^{n,n+1})}\right\rangle,
\end{eqnarray}
so that
\begin{eqnarray}
\Phi
= \frac{L}{4}\left[(c-a)\ln\frac{c}{a}+(p-q)\ln\frac{p}{q}\right]
- \frac{L}{2}\left[(c+a)\ln\frac{c}{a}\right]m
\end{eqnarray}
\begin{eqnarray}
 + \frac{L}{4}\left[(c-a)\ln\frac{c}{a} -(p-q)\ln\frac{p}{q}\right] r,
\label{Phi_many}
\end{eqnarray}
where the translational invariance was invoked.

This equation for $\Phi$ is valid at any time as long as the constraint $a+c=p+q$ is fulfilled. In the stationary state we use the result (\ref{mag}) and $r=m^2$ to get, after some algebraic manipulations, the result
\begin{eqnarray}
\nonumber
\Pi = L\frac{\sqrt{ca}}{(\sqrt{c}+\sqrt{a})^2}(p-q)\ln\frac{p}{q}.
\end{eqnarray}
When $a=c$ the entropy production rate per particle coincides with the entropy production rate obtained from the one-particle case (\ref{Phi_P}).

The current of particles $J$ is given by $pP(+,-)-qP(-+)$. But from the stationary solution (\ref{17}) we get $P(+,-)=P(-+)=P(+)P(-)=\rho(1-\rho)$ so that,  $J=\rho(1-\rho)(p-q)$, or, using (\ref{dens}),
\begin{eqnarray}
J = \frac{\sqrt{ca}}{(\sqrt{c}+\sqrt{a})^2}(p-q),
\end{eqnarray}
and we may write the entropy production rate as
\begin{eqnarray}
\Pi = L J X,
\end{eqnarray}
where $X=\ln({p}/{q})$.

%----------------------------------------------------------
\section{Pair approximation}

In this section, we will consider again the general problem of one-dimensional diffusion (with creation and annihilation of particles) where the restriction (\ref{acpq}) is removed, which means that the transition rate is
\begin{eqnarray}
w_{n}(s) = A + BS_{n} + Cs_{n+1} + Ds_{n}s_{n+1}.
\label{ABCD}
\end{eqnarray}
The dynamics of the site magnetisation $m(t):=\langle s_{i}(t)\rangle$ and pair correlation $r:=\langle s_{n}(t)s_{n+1}(t)\rangle$ in this model will be analysed through the pair approximation, and we will assume translational invariance. Multiplying (\ref{me_ff0}) by $s_{n}$ and taking the trace yields
\begin{eqnarray}
\frac{\textnormal{d}}{\textnormal{d}t}m = -2\left(B+C\right) - 4\left(A+D\right)m - 2\left(B+C\right)r.
\label{dmdt}
\end{eqnarray}
Multiplying now (\ref{me_ff0}) by $s_{n}s_{n+1}$, and taking the trace yields
\begin{eqnarray}
\frac{\textnormal{d}}{\textnormal{d}t}r = -4Ar - 2\left(B+C\right)m - 2\left(B+C\right)r_{123} - 4Dr_{13},
\label{drdt}
\end{eqnarray}
where $r_{123}:=\langle s_{n-1}s_{n}s_{n+1}\rangle=\langle s_{n}s_{n+1}s_{n+2}\rangle$ and $r_{13}:=\langle s_{n-1}s_{n+1}\rangle=\langle s_{n}s_{n+2}\rangle$. In order to close the equations (\ref{dmdt}) and (\ref{drdt}), we will express both $r_{123}$ and $r_{13}$ as functions of $m$ and $r$ following the prescription of pair approximation\cite{MT68, TO89}. Let us consider, without lack of generality, three consecutive spins, $s_{1}$, $s_{2}$ and $s_{3}$, and the marginal distribution $P(s_{1},s_{2},s_{3})$, which is obtained by summing $P(s_{1},\ldots,s_{L})$ over all the other spins. The pair approximation assumes that
\begin{eqnarray}
\nonumber P(s_{1},s_{2},s_{3}) &= P(s_{1},s_{3}|s_{2})P(s_{2}) \approx P(s_{1}|s_{2})P(s_{3}|s_{2})P(s_{2}) \\
 &= \frac{P(s_{1},s_{2})P(s_{2},s_{3})}{P(s_{2})},
\end{eqnarray}
where $P(x|y)$ stands for the conditional probability of an event $x$, given $y$. Furthermore, since
\begin{eqnarray}
P(s_{2}) = \frac{1}{2}\left(1+ms_{2}\right)
\end{eqnarray}
and
\begin{eqnarray}
P(s_{1},s_{3}) = \frac{1}{4}\left[ 1+m\left(s_{1}+s_{3}\right)+rs_{1}s_{3} \right],
\end{eqnarray}
one has
\begin{eqnarray}
\nonumber r_{123} &= \sum_{s_{1},s_{2},s_{3}}s_{1}s_{2}s_{3}P(s_{1},s_{2},s_{3}) \approx \sum_{s_{1},s_{2},s_{3}}s_{1}s_{2}s_{3}\frac{P(s_{1},s_{2})P(s_{2},s_{3})}{P(s_{2})} \\
 &= \frac{2m\left(2r-m^{2}-r^{2}\right)}{1-m^{2}}
\end{eqnarray}
and
\begin{eqnarray}
\nonumber r_{13} &= \sum_{s_{1},s_{2},s_{3}}s_{1}s_{3}P(s_{1},s_{2},s_{3}) \approx \sum_{s_{1},s_{2},s_{3}}s_{1}s_{3}\frac{P(s_{1},s_{2})P(s_{2},s_{3})}{P(s_{2})} \\
 &= \frac{2\left(m^{2}-2m^{2}r+r^{2}\right)}{1-m^{2}}.
\end{eqnarray}
The set of differential equations (\ref{dmdt}) and (\ref{drdt}) in the pair approximation becomes, therefore,
\begin{eqnarray}
\left\{
\begin{array}{ll}
\displaystyle\frac{\textnormal{d}}{\textnormal{d}t}m &= \displaystyle\left(c-a\right) - 2\left(c+a\right)m + \left(c-a\right)r \\
 & \\
\displaystyle\frac{\textnormal{d}}{\textnormal{d}t}r &= \displaystyle\left(c-a\right)m + 2\left(c-a\right)\frac{m\left(2r-m^{2}-r^{2}\right)}{1-m^{2}} + \\
 &\displaystyle + 4\left(p+q-c-a\right)\frac{m^{2}-2m^{2}r+r^{2}}{1-m^{2}} -\left(p+q+c+a\right)r
\end{array}
\right.
\label{system2pair_2}
\end{eqnarray}
in the original parameters $p$, $q$, $c$ and $a$, introduced in section \ref{problem}. The entropy flux per site is given by
\begin{eqnarray}
\nonumber \frac{\Phi(t)}{L} &= \frac{1}{4}\left[\left(p-q\right)\ln\frac{p}{q} + \left(c-a\right)\ln\frac{c}{a}\right] - m(t)\left(\frac{c+a}{2}\right)\ln\frac{c}{a} - \\
 &- \frac{1}{4}\left[\left(p-q\right)\ln\frac{p}{q} - \left(c-a\right)\ln\frac{c}{a}\right]r(t),
\label{2pair_flux}
\end{eqnarray}
and its time evolution can be obtained numerically from the system (\ref{system2pair_2}) above, as shown in Figure 2.

%-------------
\begin{center}
\begin{figure}
\begin{center}
\includegraphics[width=210pt, height=140pt]{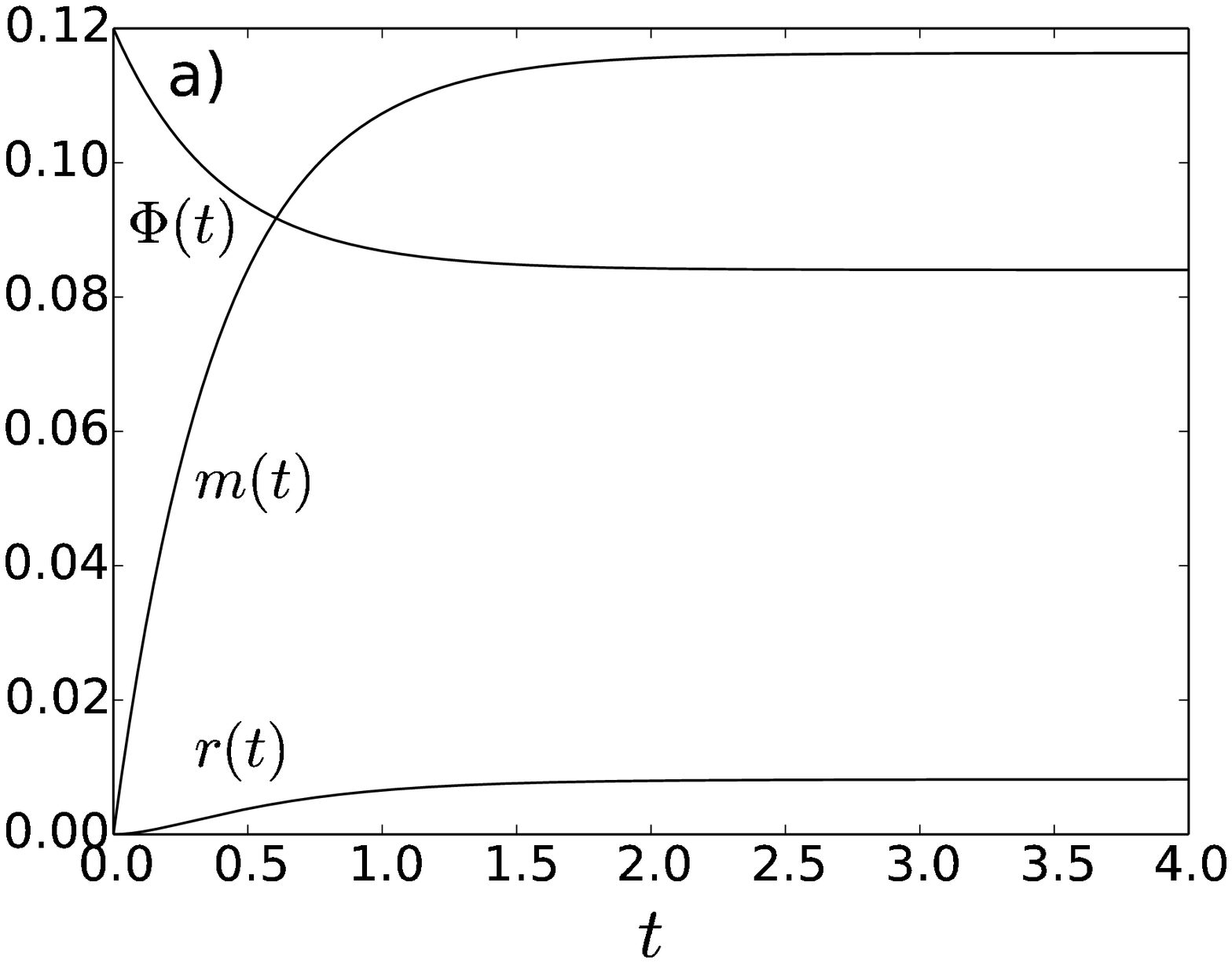}
\includegraphics[width=210pt, height=140pt]{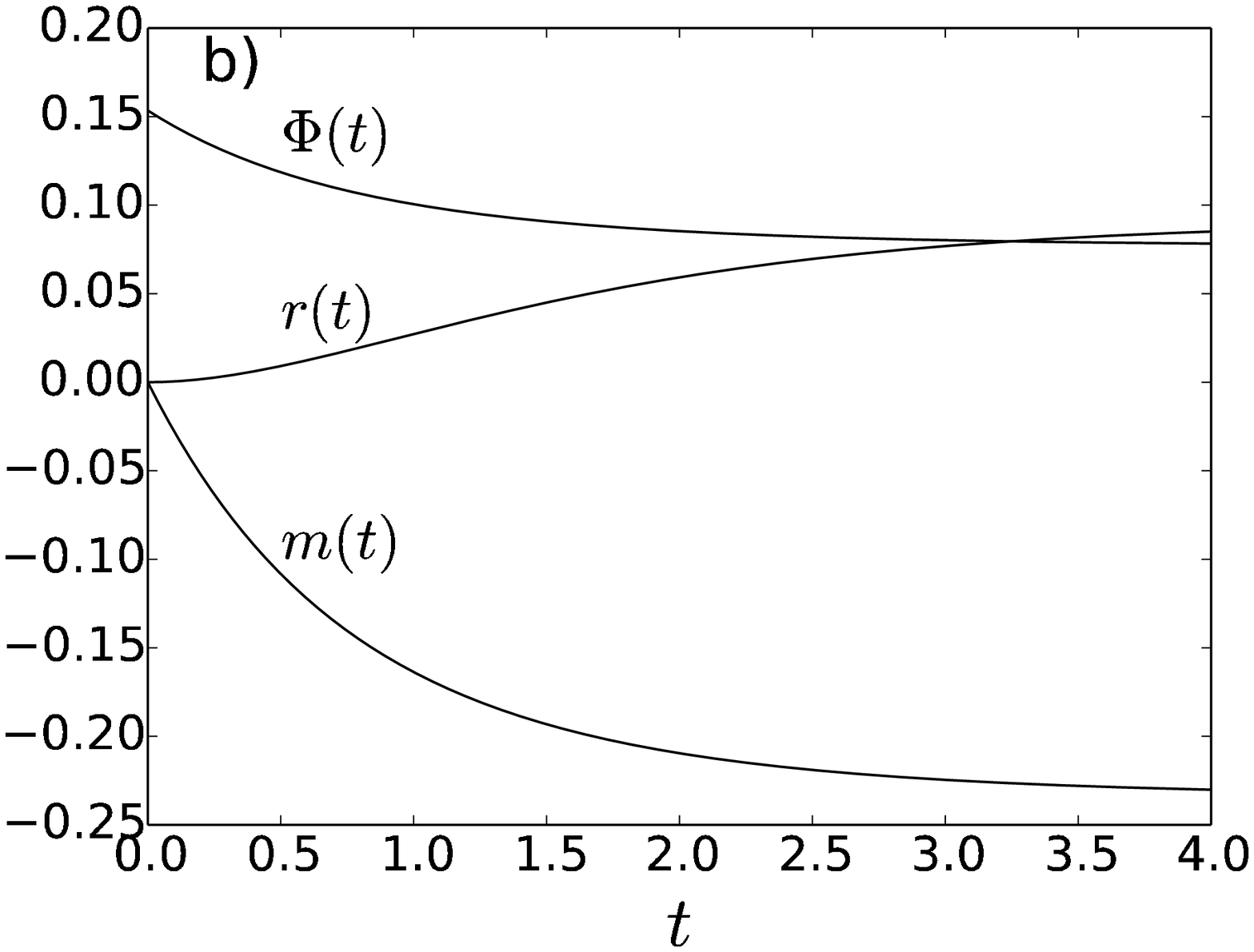}
\\
\includegraphics[width=210pt, height=140pt]{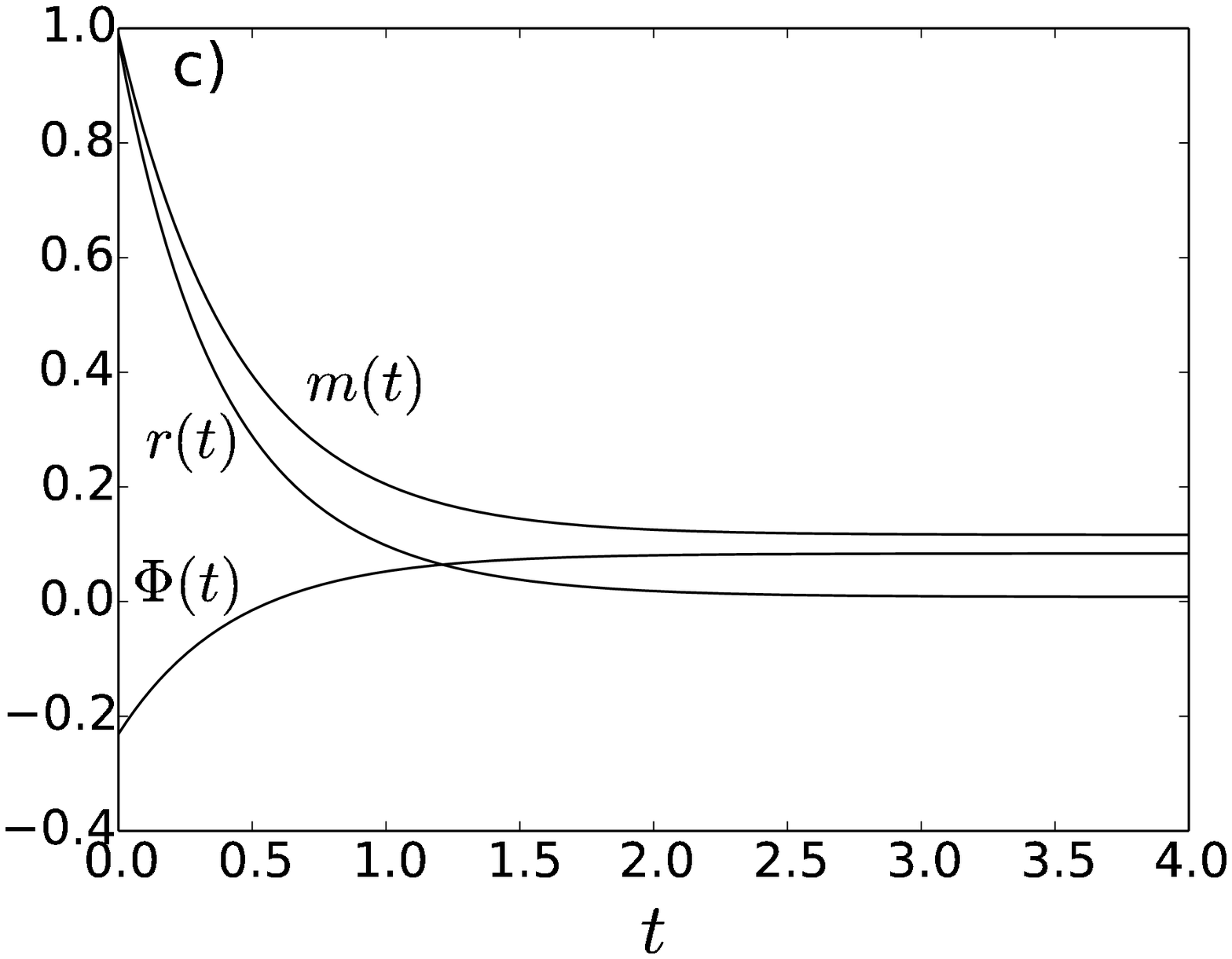}
\includegraphics[width=210pt, height=140pt]{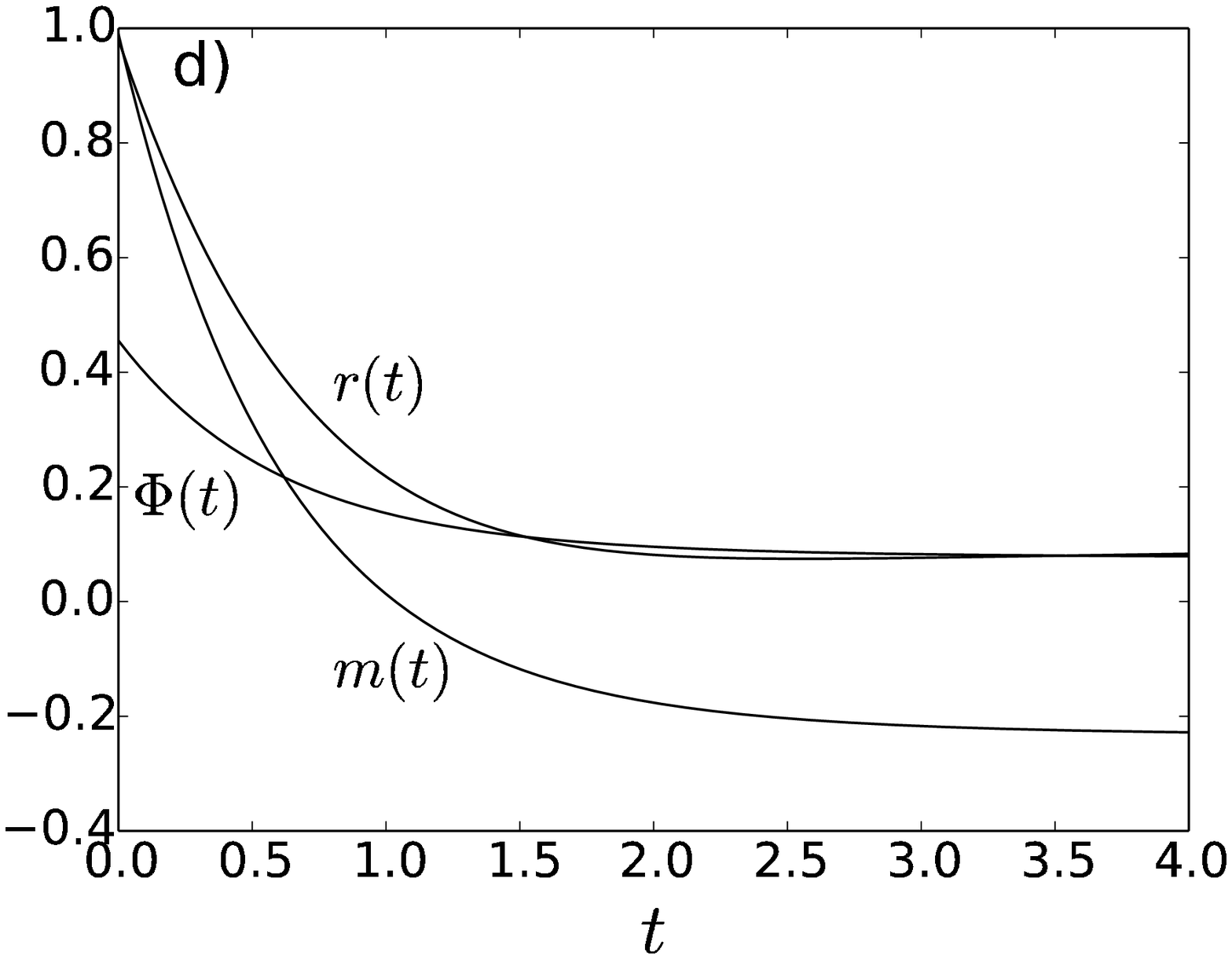}
\end{center}
\caption{Graphs of $m(t)$, $r(t)$ and $\Phi(t)$ with time in pair approximation with $p=0.7$ and $q=0.3$. 2a) $c=0.8$, $a=0.5$ and initial condition $m(0)=r(0)=0.0$. 2b) $c=0.2$, $a=0.5$ and initial condition $m(0)=r(0)=0.0$. 2c) $c=0.8$, $a=0.5$ and initial condition $m(0)=0.99$ and $r(0)=0.99^{2}$. 2d) $c=0.2$, $a=0.5$ and initial condition $m(0)=0.99$ and $r(0)=0.99^{2}$.}
\end{figure}
\end{center}
%-----------

Nevertheless, some analytical results are available from the equations above in the stationary regime. Let us denote by $m_{\infty}$ and $r_{\infty}$ the stationary magnetisation and pair correlation, respectively. If $a=c=\frac{p+q}{2}$, the stationary solution is trivial, $m_{\infty}=r_{\infty}=0$. On the other hand, if $a=c$ with $a\neq\frac{p+q}{2}$, then $m_{\infty}=0$, but there are two possibilities for $r_{\infty}$: besides the trivial solution $r_{\infty}=0$, there is another one, which is
\begin{eqnarray}
r_{\infty} = \frac{p+q+2c}{4\left(p+q-2c\right)} \qquad (a=c\neq\frac{p+q}{2}).
\end{eqnarray}
The stability analysis shows that the point $(m_{\infty},r_{\infty})=(0,0)$ is a stable point, while $(m_{\infty},r_{\infty})=(0,\frac{p+q+2c}{4\left(p+q-2c\right)})$ is a saddle point.

When $a=c$, the entropy flux can be written as
\begin{eqnarray}
\frac{\Phi(\infty)}{L} = \frac{1}{4}\left(1-r_{\infty}\right)\left(p-q\right)\ln\frac{p}{q} \qquad (a=c),
\label{2pair_flux_st_a=c}
\end{eqnarray}
and the entropy production has the same value, $\Pi(\infty)=\Phi(\infty)$ in the stationary state.

Finally, the condition $a\neq c$ leads to
\begin{eqnarray}
\left\{
\begin{array}{ll}
\displaystyle r_{\infty}-2\left(\frac{c+a}{c-a}\right)m_{\infty}+1 &=0 \\
 & \\
\displaystyle m_{\infty} + \frac{2m_{\infty}\left(2r_{\infty}-m_{\infty}^{2}-r_{\infty}^{2}\right)}{1-m_{\infty}^{2}} - \left(\frac{c+a}{c-a}\right)r_{\infty} &= 0
\end{array}
\right.,
\end{eqnarray}
and one should solve a polynomial equation of third order to obtain an analytical result. The entropy flux, which is equal to the entropy production, is
\begin{eqnarray}
\frac{\Phi(\infty)}{L} = \frac{1}{2}\left[1-\left(\frac{c+a}{c-a}\right)m_{\infty}\right]\left(p-q\right)\ln\frac{p}{q}.
\label{2pair_flux_st_aneqc}
\end{eqnarray}
Note that the behaviour of the magnetisation, pair correlation and entropy are invariant under the exchange $p\leftrightarrow q$. Moreover, the form
\begin{eqnarray}
\Pi(\infty) \propto L\left(p-q\right)\ln\frac{p}{q}
\end{eqnarray}
in the pair approximation also recovers the dependence on $p$ and $q$ seen in the previous section for the entropy production.

%----------------------------------------------------------
\section{Discussion}

The dynamics of an asymmetric exclusion process was considered on a ring. In the case where the dynamics was restricted to a single particle, the entropy flux $\Phi$, which turned to be a constant, was determined. This constant is zero in the symmetric case, as expected. The entropy production, on the other hand, was evaluated in the continuous approximation, and it was shown that it decays algebraically to $\Phi$, which is again constant during the whole dynamics. In the case of many particles with the constraint $a+c=p+q$, we have determined the stationary probability in a closed form from which we have found the entropy production rate. The general many-particles case, where this constraint is absent, was treated at pair approximation level, and we could characterise numerically the time-dependent behaviour of the entropy flux and determined analytically its stationary value, which coincides with the stationary entropy production. It is worth mentioning that although the stationary distribution has the Boltzmann-Gibbs form, the system is not thermodynamic equilibrium if $p\neq q$. From the thermodynamic point of view, the asymmetric exclusion process should thus be considered a non-equilibrium system. Finally, in all models studied here, the entropy production rate, in the stationary state, has a bilinear form in the current $J$ and the force $X$.

%----------------------------------------------------------
%\section{Acknowledgements}

%----------------------------------------------------------


\section*{References}

\begin{thebibliography}{99}

\bibitem{NP77} G. Nicolis and I. Prigogine I, 
{\it Self-Organization in Nonequilibrium Systems},
Wiley, New York, 1977.

\bibitem{vK81} N. G. van Kampen N G,
{\it Stochastic Processes in Physics and Chemistry},
North-Holland, Amsterdam, 1981.

\bibitem{TO15} T. Tom\'e T and M. J. de Oliveira.
{\it Stochastic Dynamics and Irreversibility}, 
Springer, Heidelberg, 2015.

\bibitem{S76} J. Schnakenberg, 
Rev. Mod. Phys. {\bf 48}, 571 (1976).

\bibitem{J84} L. Jiu-Li, C. Van den Broeck and G. Nicolis,
Z. Phys. B {\bf 56},165 (1984).

\bibitem{M86} C. Y. Mou, J.-L. Luo and G. Nicolis,
J. Chem. Phys. {\bf 84}, 7011 (1986).

\bibitem{CT05} L. Crochik and T. Tom\'e, Phys. Rev. E
{\bf 72}, 057103 (2005). 

\bibitem{Z06} R. K. P. Zia and B Schmittmann,
J. Phys. A: Math. Gen. {\bf 39}, L407 (2006).

\bibitem{A06} D. Andrieux and P. Gaspard,
Phys. Rev. E {\bf 74}, 011906 (2006).

\bibitem{S07} T. Schmiedl and U. Seifert,
J. Chem. Phys. {\bf 126}, 044101 (2007).

\bibitem{Z07} R. K. P. Zia and B. Schmittmann,
J. Stat. Mech. P07012 (2007).

\bibitem{E09} M. Esposito, K. Lindenberg, and
C. Van den Broeck, Phys. Rev. Lett. {\bf 102}, 130602 (2009).

\bibitem{TO10} T. Tom\'e and M. J. de Oliveira,
Phys. Rev. E {\bf 82}, 021120 (2010).

\bibitem{O11} M. J. de Oliveira, 
J. Stat. Mech. P12012 (2011).

\bibitem{E12} M. Esposito, 
Phys. Rev. E {\bf 85}, 041125 (2012).

\bibitem{HO12} M. O. Hase and M. J. de Oliveira,
J. Phys. A {\bf 45}, 165003 (2012).

\bibitem{TO12} T. Tom\'e and M. J. de Oliveira,
Phys. Rev. Lett. {\bf 108}, 020601 (2012). 

\bibitem{T06} T. Tom\'e,
Braz. J. Phys. {\bf 36}, 1285 (2006).

\bibitem{MO98} J. R. G. de Mendon\c{c}a and M. J. de Oliveira,
J. Stat. Phys. {\bf 92}, 651 (1998).

\bibitem{O99} M. J. de Oliveira,
Phys. Rev. E {\bf 60}, 2563 (1999).

\bibitem{MT68} H. Mamada and F. Takano,
J. Phys. Soc. Jap. {\bf 25}, 675 (1968).

\bibitem{TO89} T. Tom\'e and M. J. de Oliveira,
Phys. Rev. A {\bf 40}, 6643 (1989).

\end{thebibliography}
\end{document}